\pdfoutput=1
\documentclass[aps,prl,floatfix,twocolumn,reprint,superscriptaddress]{revtex4}
\usepackage{amsfonts}
\usepackage{mathrsfs}
\usepackage{amsmath}
\usepackage{color}
\usepackage{graphicx}
\usepackage{bm}
\usepackage{amssymb}
\usepackage{xspace}
\usepackage{epstopdf}
\usepackage{dcolumn}
\usepackage{longtable}
\usepackage[colorlinks=true, letterpaper=true, pdfstartview=FitV, linkcolor=blue, citecolor=blue, urlcolor=blue]{hyperref}

\begin{document}
\title{ Weak Topological Insulators and Composite Weyl Semimetals: $\bm\beta$-Bi$_{\bm 4}$X$_{\bm 4}$ (X=Br,~I)
}
\author{Cheng-Cheng Liu}
\affiliation{Department of Physics, University of Texas at Dallas, Richardson, Texas 75080, USA}
\affiliation{School of Physics, Beijing Institute of Technology, Beijing 100081, China}
\author{Jin-Jian Zhou}
\affiliation{School of Physics, Beijing Institute of Technology, Beijing 100081, China}
\author{Yugui Yao}\email{ygyao@bit.edu.cn}
\affiliation{School of Physics, Beijing Institute of Technology, Beijing 100081, China}
\author{Fan Zhang}\email{zhang@utdallas.edu}
\affiliation{Department of Physics, University of Texas at Dallas, Richardson, Texas 75080, USA}

\begin{abstract}
While strong topological insulators (STI) have been experimentally realized soon after their theoretical predictions,
a weak topological insulator (WTI) has yet to be unambiguously confirmed. 
A major obstacle is the lack of distinct natural cleavage surfaces to test the surface selective hallmark of WTI.
With a new scheme, we discover that Bi$_{4}$X$_{4}$ (X=Br,~I), stable or synthesized before, 
can be WTI with two natural cleavage surfaces, where two anisotropic Dirac cones stabilize and annihilate, respectively. 
We further find four surface state Lifshitz transitions under charge doping and two bulk topological phase transitions under uniaxial strain.
Near the WTI-STI transition, there emerges a novel Weyl semimetal phase,
in which the Fermi arcs generically appear at both cleavage surfaces 
whereas the Fermi circle only appears at one selected surface. 
\end{abstract}
\maketitle

The discovery of topological insulators~\cite{Hasan-10,Qi-11,Moore-10} has led to an ongoing revolution deepening our fundamental understanding of quantum materials.
The controllable bulk topological quantum phase transitions and the protected spin-momentum locked surface states
may ultimately lead to unprecedented advances in technologies, e.g., the Majorana-based fault-tolerant quantum computing~\cite{Stern-15}. 
Historically, strong topological insulators (STI) and weak topological insulators (WTI) were predicted together to exist in three dimensions~\cite{Fu-07,Moore-07,Roy-09}.
Astonishingly, STI are not at all rare in nature. Over a dozen documented materials have been identified as STI under readily accessible experimental conditions~\cite{Xia-09,Chen-09,Franz-10,Yang-12,Mele}.
Notably, their lateral structure offers a superior advantage to observe the hallmark of STI, a surface Dirac cone, without intended surface passivation that could be challenging.
However, the unambiguous experimental confirmation of WTI is still elusive~\cite{Mele}.
Thus far there have been two routes to construct WTI, i.e., to stack weakly coupled layers of quantum spin Hall insulators (QSHI)~\cite{Yan-12,Brink-13},
or to engineer a superlattice of alternating layers with multiple band inversions~\cite{Liu-14,Feng-14}.
Evidently, these designer WTI  pose extreme experimental challenges.
In the first route, the protected surface metallicity will only be present at non-cleavage surfaces (non-parallel to the layers), yet surface roughness and dangling bonds would prevent us from observing this hallmark~\cite{Brink-13}. 
In the second route, though the surface states can survive at the cleavage surface (parallel to the layers), the intra- and inter-layer couplings within a supercell must be fine tuned in material synthesis.
Therefore, to unambiguously determine the existence of WTI and to explore the exotic phenomena uniquely hosted by WTI,
it is crucial to develop a new route. 

We propose to realize the WTI in a van der Waals (vdW) material that is a periodic stack of one dimensional atomic chains.
Such a WTI possesses two natural cleavage planes, enabling respective observations of stabilization and annihilation of distinct surface Dirac cones.
We discover that $\beta$-Bi$_{4}$X$_{4}$ (X=I,~Br) can be such a class of WTI.
Markedly, with a chemical formula as simple as the prototype STI $\rm Bi_2X_3$ (X=Se,~Te), 
$\beta$-Bi$_{4}$X$_{4}$ (X=I,~Br) are real crystalline solids rather than artificial periodic heterostructures.
$\beta$-Bi$_{4}$I$_{4}$ has been successfully grown as a large crystal before~\cite{Benda1978-1,Benda1978-2,Dikarev2001,Filatova2007}, 
and $\beta$-Bi$_{4}$Br$_{4}$ is demonstrated to be similarly stable~\cite{SM}. 
After revealing their unique crystal structure, topological band properties, and surface Lifshitz transitions,
we will further examine their topological phase transitions under uniaxial strain. 
Intriguingly, a novel Weyl semimetal (WSM) phase emerges near the WTI-STI transition;
in addition to the existence of Fermi arcs at both cleavage surfaces as the case of WSM,
a Fermi circle only exists at one selected surface reminiscent of the case of WTI. 
As prototype WTI, $\beta$-Bi$_{4}$X$_{4}$ (X=I,~Br) offers a new platform for exploring exotic physics with simple chemistry.

\begin{figure}[b!]
\includegraphics[width=0.76\columnwidth]{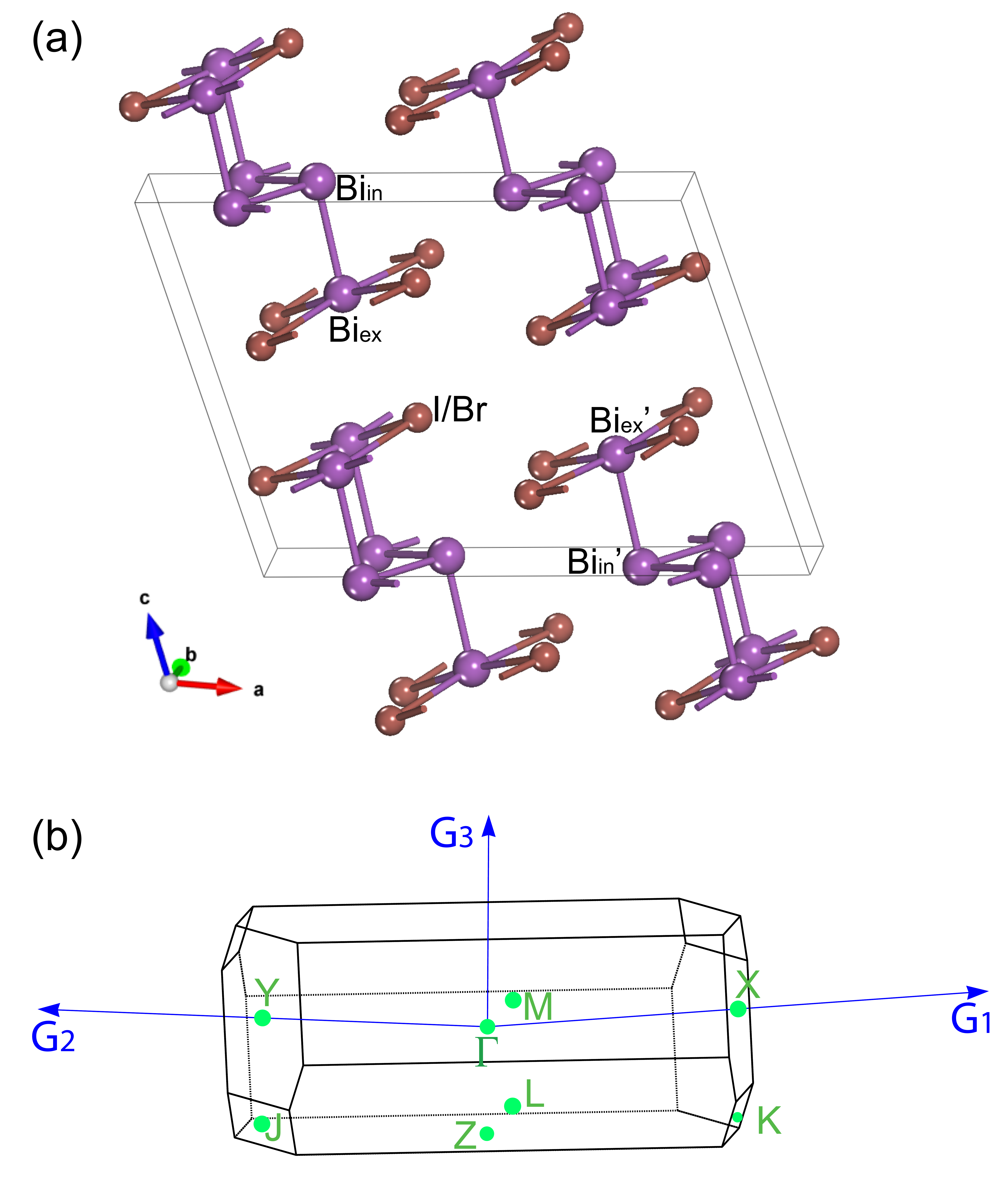}
\caption{{\bf Crystal structure and Brillouin zone of $\bm\beta$-Bi$_{\bm 4}$X$_{\bm 4}$ (X=I,~Br).} 
(a) The crystal structure of the conventional cell of $\beta$-Bi$_4$X$_4$.  
Bi$_{\rm in}$ (Bi$_{\rm ex}$) and Bi$_{\rm in}'$ (Bi$_{\rm ex}'$) atoms are related by inversion symmetry. 
(b) The first Brillouin Zone of $\beta$-Bi$_4$X$_4$ with eight time-reversal-invariant momenta.
$\bm G$ are the reciprocal lattice vectors.}\label{fig1}
\end{figure}

\begin{figure*}[t!]
\includegraphics[width=1.9\columnwidth]{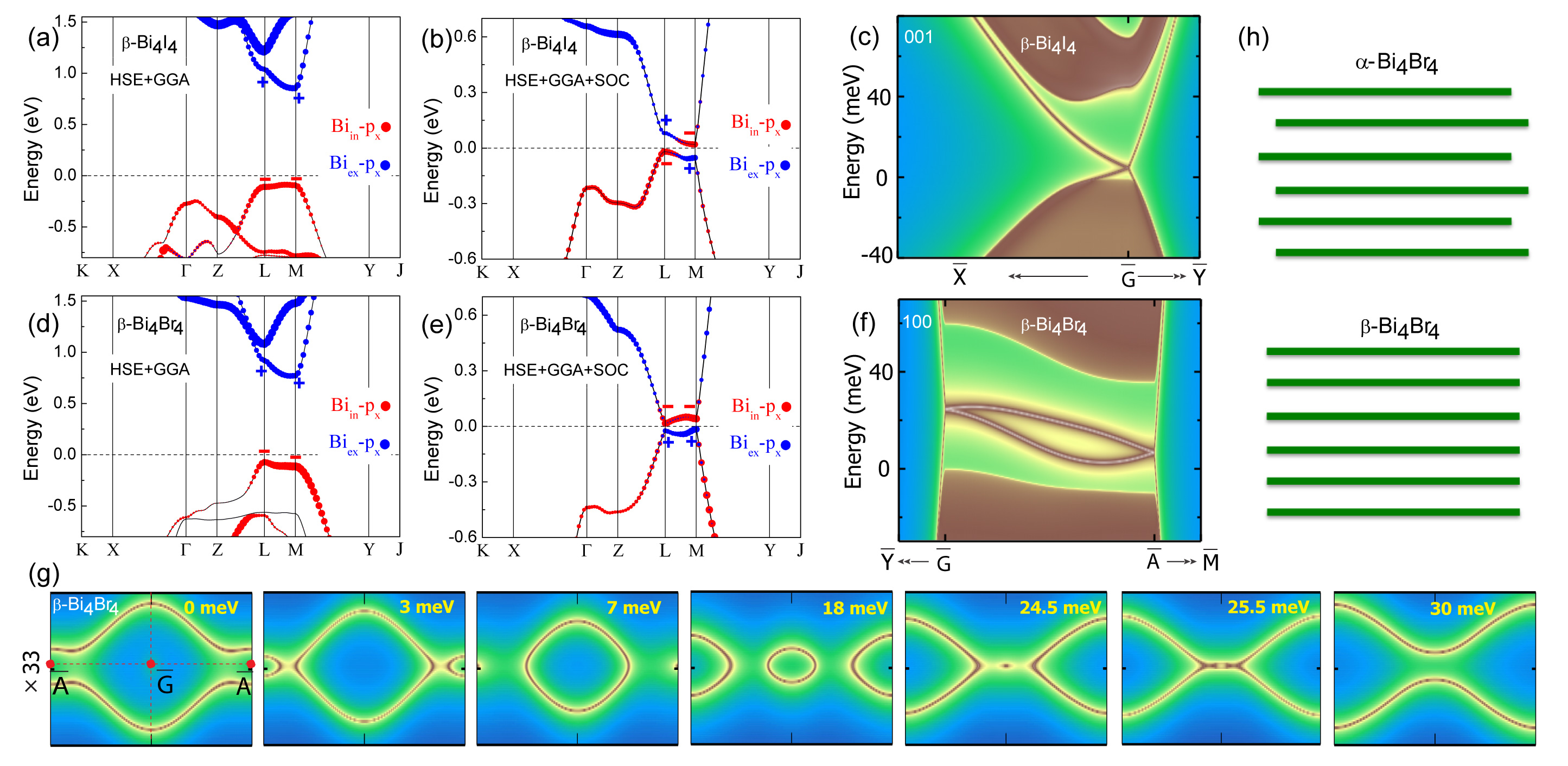}
\caption{{\bf Band structures and surface states of $\bm\beta$-Bi$_{\bm 4}$X$_{\bm 4}$ (X=I,~Br).}
(a)-(b) The bulk band structures for $\beta$-Bi$_4$I$_4$ without and with the SOC.
The size of red (blue) dots indicates the weight of the relevant $p_x$ orbital of Bi$_{\rm in}$ (Bi$_{\rm ex}$) atoms;
the symbols $\pm$ label the parities of the bands at $L$ and $M$ points;
the dashed lines are the Fermi levels. 
(c) The $(001)$ cleave-surface states of $\beta$-Bi$_4$I$_4$.
(d)-(e) The same as (a)-(b) but for $\beta$-Bi$_4$Br$_4$.
(f) The $(100)$ cleave-surface states of $\beta$-Bi$_4$Br$_4$ and (g) their Lifshitz transitions. 
Vertical dimensions in (g) are magnified by $33$ times for clarity.
(h) The illustration of the stacking orders in $\alpha$- and $\beta$-Bi$_{\bm 4}$Br$_{\bm 4}$.
The $\beta$ ($\alpha$) phase is a stack of single (double) $(001)$ layers; each layer denoted by a green line is a QSHI.}
\label{fig2}
\end{figure*} 

\noindent{\bf Topological Band Structures}

Both Bi$_4$I$_4$ and Bi$_4$Br$_4$ have stable $\alpha$ and $\beta$ phases. The two phases crystallize in the same monoclinic space group $C_{2h}^3$ ($C2/m$), 
and differ only in the way their building blocks are stacked~\cite{Benda1978-1,Benda1978-2,Dikarev2001,Filatova2007}.  
The $\alpha$ phases turn out to be trivial insulators~\cite{SM}.
As the $\beta$ phases have rather similar lattice structures, 
by way of illustration we refer to $\beta$-Bi$_4$I$_4$ that has been experimentally synthesized,
and the explicit lattice constants and atom positions of $\beta$-Bi$_4$Br$_4$ are given in Table S1~\cite{SM}.
Figure~\ref{fig1}a shows the conventional cell of $\beta$-Bi$_4$I$_4$, in which $a=14.386$~\AA, $b=4.430$~\AA, $c=10.493$~\AA, and $\beta=107.87^\circ$.
One unit cell consists of four I atoms and four Bi atoms that can be divided into two types.
The two internal Bi atoms, labeled as Bi$_{\rm in}$ and Bi$_{\rm in}'$, form zigzag atomic chains with the nearest neighbor distance of $3.04$~\AA.
The two external Bi atoms, labeled as Bi$_{\rm ex}$ and Bi$_{\rm ex}'$, 
are each bonded to four I atoms with the Bi-I distance of $3.14$~\AA\ and to one internal Bi atom with a distance of $3.06$~\AA. 
The crystal has two independent symmetries, spatial inversion and mirror reflection.
Under $(010)$ mirror reflection, the four Bi atoms are invariant.
Upon spatial inversion, the two Bi atoms are interchanged for each type. 
As suggested by Fig.~\ref{fig1}a, $\beta$-Bi$_4$I$_4$ is a periodic stack of atomic chains aligned to the $b$ direction.
Indeed, we find that the interlayer binding energies for $(100)$ and $(001)$ planes are about $20$~meV/\AA$^2$,
comparable to $12$~meV/\AA$^2$ for graphite~\cite{Liu2012} and $26$~meV/\AA$^2$ for MoS$_2$~\cite{Zhou2014}.
Evidently, $\beta$-Bi$_4$X$_4$ (X=I,~Br) are typical van der Waals (vdW) materials, but with {\it two} cleavage surfaces. 

We employ the HSE hybrid functional method~\cite{CM}, which is believed to be more sophisticated and accurate than the GGA and LDA methods, 
to carry out our density functional theory (DFT) calculations of the electronic band structures of $\beta$-Bi$_4$I$_4$ and $\beta$-Bi$_4$Br$_4$.
The vdW corrections~\cite{CM} and the lattice relaxations are also taken into account to optimize the crystal structures.
The projected band structures of $\beta$-Bi$_4$I$_4$ without and with SOC are shown in Figs.~\ref{fig2}a-\ref{fig2}b, respectively. 
The bands near the gap are mainly contributed from the Bi $p_x$ orbitals. 
Without the SOC, the conduction and valence band edges at both $L$ and $M$ points are respectively from the Bi$_{\rm ex}$ and Bi$_{\rm in}$ orbitals, 
which are identified to exhibit opposite parities.
When the SOC is included, the constituents and parities of the conduction and valence bands remain the same at the $L$ point, whereas they are inverted at the $M$ point.
Based on the Fu-Kane parity criterion~\cite{FuParity}, we can conclude that $\beta$-Bi$_4$I$_4$ is a STI, with $(1;110)$ $\mathcal{Z}_2$ invariants and a $39$~meV indirect gap.  
As plotted in Fig.~\ref{fig2}c, we further obtain one surface Dirac cone at the $(001)$ cleavage surface, which is consistent with the conclusion that $\beta$-Bi$_4$I$_4$ is a STI.
Notably, the Dirac cone is highly anisotropic, because the crystal only has inversion and reflection symmetries, and because the surface is parallel to the atomic chains.

We now study the band structure of $\beta$-Bi$_4$Br$_4$, which share evident similarities to the case of $\beta$-Bi$_4$I$_4$.
However, the band inversion occurs at both $M$ and $L$ points in $\beta$-Bi$_4$Br$_4$, as shown in Figs.~\ref{fig2}d-\ref{fig2}e.
It follows from the Fu-Kane parity criterion~\cite{FuParity} that $\beta$-Bi$_4$Br$_4$ is a WTI with $(0;001)$ $\mathcal{Z}_2$ invariants and a $32$~meV indirect gap.
We further calculate the surface states for the two natural cleave surfaces, i.e., the $(100)$ and $(001)$ surfaces. 
Since $M$ and $L$ are projected into the same point at the $(001)$ surface Brillouin zone, 
the $(001)$ surface are anticipated to host neither protected surface states nor dangling bond states. Our calculations verify this physical picture (not shown).
In contrast, at the $(100)$ surface, $M$ and $L$ are projected into two distinct points $\bar{G}$ and $\bar{A}$.
As anticipated and shown in Fig.~\ref{fig2}f, our calculations identify two (connected) surface Dirac cones at $\bar{G}$ and $\bar{A}$ points; 
the dangling bond states are completely absent.

The $(100)$ surface states of $\beta$-Bi$_4$Br$_4$ also exhibit prominent anisotropy, for the reasons that we have discussed above.
The direction with the much larger Fermi velocity is along $\bm b$, in which the atomic chains are aligned.
Moreover, the two surface Dirac points at $\bar{G}$ and $\bar{A}$ are at different energies.
These two features enrich the Fermi surface topology of the $(100)$ surface states. 
Fig.~\ref{fig2}g characterizes the four corresponding Lifshitz transitions, as we now explain in ascending energy.
At zero energy, the two hole pockets around $\bar{G}$ and $\bar{A}$ are connected. 
As the energy increases, e.g., to $3$~meV, the two pockets are disconnected after the first Lifshitz transition.
At $7$~meV, the second transition occurs; the hole pocket at $\bar{A}$ contracts into a Dirac point, followed by the emergence of an electron pocket. 
At higher energies, the disconnected electron and hole pockets coexist up to the third transition at about $24$~meV, 
in which the hole pocket at $\bar{G}$ contracts into a Dirac point and then an electron pocket emerges.
Ultimately, the fourth transition occurs and the two electron pockets become connected.

We note by passing that, in sharp contrast to the case of $\beta$ phase, both $\alpha$-Bi$_4$I$_4$ and $\alpha$-Bi$_4$Br$_4$ are normal insulators (NI)~\cite{SM}. 
The contrasting topological properties of the $\alpha$ and $\beta$ phases can be understood in the way illustrated by Fig.~\ref{fig2}h.
The $\beta$ phase is a stack of single $(001)$ layers, whereas the $\alpha$ phase is a stack of double $(001)$ layers.
For Bi$_4$Br$_4$, the single $(001)$ layer was demonstrated to be a QSHI with a $0.18$~eV gap~\cite{Zhou2014}.
Given the weak interlayer couplings, the $\beta$ phase is a WTI with similar band inversions at $k_z=0$ and $k_z=\pi$ planes, 
whereas the $\alpha$ phase is trivial since each bilayer is a NI.
For Bi$_4$I$_4$, the single $(001)$ layer was shown to be close to the QSHI-NI critical point~\cite{Zhou2014}.
Thus, the $\alpha$ phase is definitely a NI, yet the $\beta$ phase is not necessarily a WTI, depending on the details.
However, as we will show, $\beta$-Bi$_4$I$_4$ can become WTI under uniaxial strain.   \\

\begin{figure}
\includegraphics[width=1.0\columnwidth]{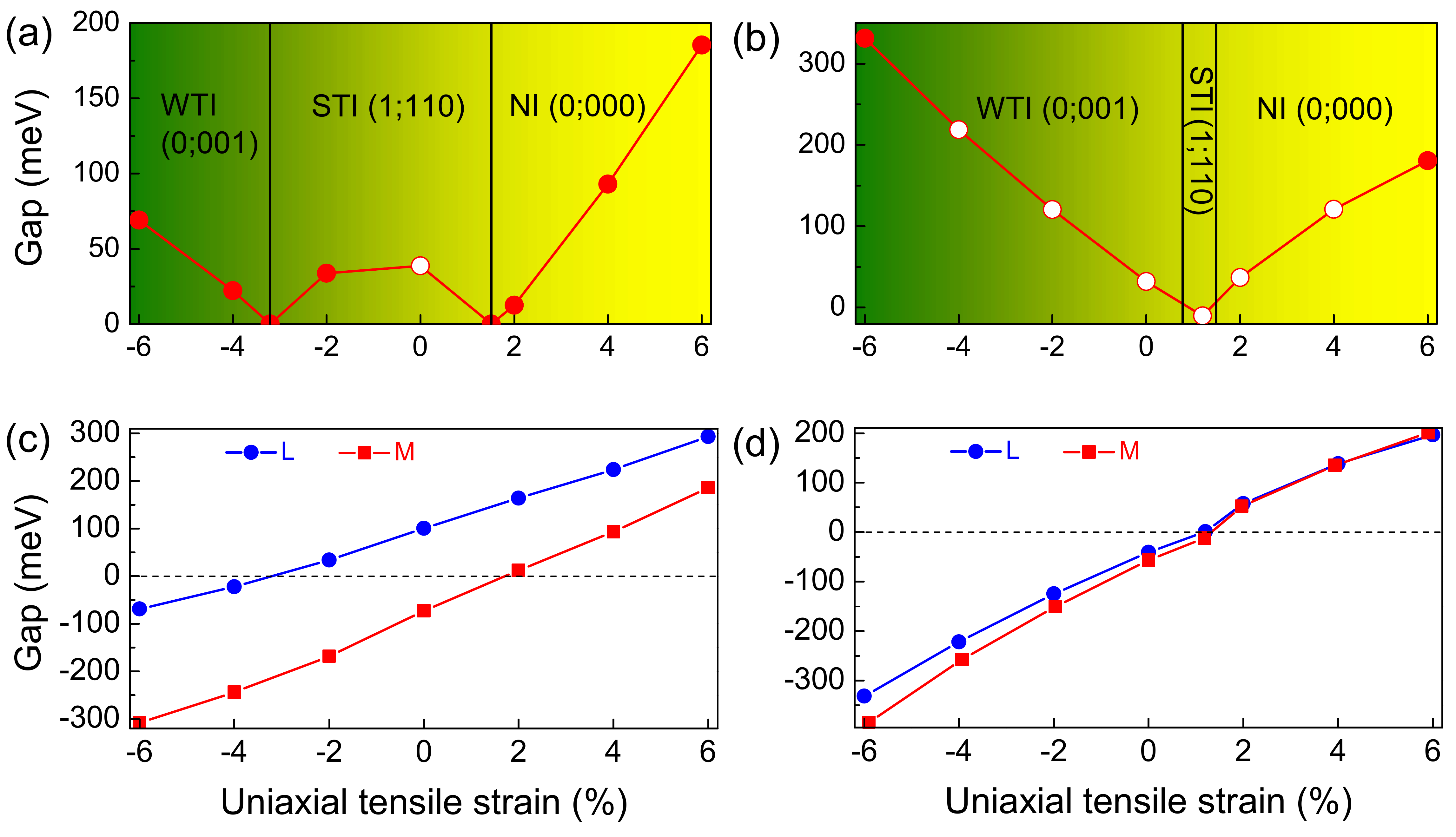}
\caption{{\bf Topological phase transitions and band inversions under uniaxial strain along the $\bm a$ axis.} 
The phase diagrams versus the uniaxial strain are depicted in (a) for  $\beta$-Bi$_4$I$_4$ and (b) for $\beta$-Bi$_4$Br$_4$. 
The corresponding direct gaps at $L$ and $M$ points are shown in (c) for $\beta$-Bi$_4$I$_4$ and (d) for $\beta$-Bi$_4$Br$_4$. 
A solid (open) dot denotes a direct (indirect) gap;
the minus sign of a gap indicates a band inversion.}
\label{fig3}
\end{figure}

\noindent{\bf Topological Phase Transitions}

We now construct an effective model for the $\beta$ phases. 
The time-reversal-invariant $M$ and $L$ points have the little group $C_{2h}^{3}$ with two independent symmetries, spatial inversion $\mathcal{P}=\tau_z$ and mirror reflection $\mathcal{M}_y=i\sigma_y$. 
Here $\bm\tau$ and $\bm \sigma$ are the orbital and spin Pauli matrices.
By convention we choose the time-reversal operator $\mathcal{T}=iK\sigma_y$ with $K$ the complex conjugation. 
Given the three symmetries, to the linear order the $k\cdot p$ Hamiltonians near $L$ and $M$ may be written as~\cite{footnote} 
\begin{eqnarray}\label{Heff}
\mathcal{H}^i = v_x^i k_x \sigma_y \tau_x + v_y^i k_y \sigma_x\tau_x + v_z^i k_z\tau_y + m^i\tau_{z} + c^i,
\end{eqnarray} 
where $i$ refers to the $M$ or $L$ point, $\bm v$ are the velocities, $c$ is the energy offset,
and $m$ is the energy gap with $m<0$ denoting an inverted gap. 
For $\beta$-Bi$_4$I$_4$ only $m^{M}$ is negative; for $\beta$-Bi$_4$Br$_4$ both $m^{M}$ and $m^{L}$ are negative.

With Eq.~(\ref{Heff}), we elaborate the strain effects on $\beta$-Bi$_4$X$_4$ (X=I,~Br).
The strain tensor $\varepsilon_{ij}$ is a rank-2 tensor and invariant under time reversal.  
With these two facts and the original symmetries of $\mathcal{H}$, the strain induced perturbations to the lowest order take the form of 
\begin{eqnarray}\label{Hstrain}
\delta\mathcal{H}^i =  \delta\mathcal{H}^i_0 + \left(\varepsilon_{11}\lambda_{11}^i+\varepsilon_{22}\lambda_{22}^i+\varepsilon_{33}\lambda_{33}^i+\varepsilon_{13}\lambda_{13}^i\right)\tau_{z},
\end{eqnarray} 
where $\lambda_{ij}$ are deformation potentials, and $\delta\mathcal{H}^{i}_0$ is a rigid shift of all bands with a form similar to that in the parentheses.
Clearly, applying strain along the $\bm a$, $\bm b$, or $\bm c$ axis would change the direct gaps and hence the band inversions at the $M$ and $L$ points.
Moreover, the direct gaps should exhibit linear dependence on the small strain, and the inversion symmetry is respected in the deformation.
It follows that the topological phase transitions among NI, STI, and WTI can be tuned by the uniaxial strain in $\beta$-Bi$_4$X$_4$ (X=I,~Br).

This prediction can be verified by our DFT calculations~\cite{CM} with strain along the $\bm a$ axis, as shown in Fig.~\ref{fig3}. 
In this case, the deformation potentials at $M$ and $L$ points exhibit the same sign and similar magnitudes.
For the STI $\beta$-Bi$_4$I$_4$, under more than $1.5\%$ tensile strain, the inverted bands at $M$ become un-inverted while those at $L$ remain not inverted, yielding a STI to NI transition. On the other hand, under more than $3.3\%$ compressive strain, the bands at $L$ become inverted while those at $M$ remain inverted, producing a STI to WTI transition. 
For the WTI $\beta$-Bi$_4$Br$_4$, the inverted gaps at $L $ and $M$ both increase with increasing the compressive strain, resulting a larger gap WTI.
As the tensile strain increases, however, there are two successive transitions from the WTI to a STI and then to a NI, 
because the bands at $L$ become un-inverted prior to those at $M$.
Figures~\ref{fig3}a-\ref{fig3}d plot the phase diagrams and the direct gaps at $L$ and $M$, 
which exhibit linear dependence on the strain predicted by our effective model.
Similar phase transitions can also be driven by strain along other axes~\cite{SM}.
These results are suggestive of a feasible way to engineer or stabilize the WTI phase.\\

\begin{figure}
\includegraphics[width=0.8\columnwidth]{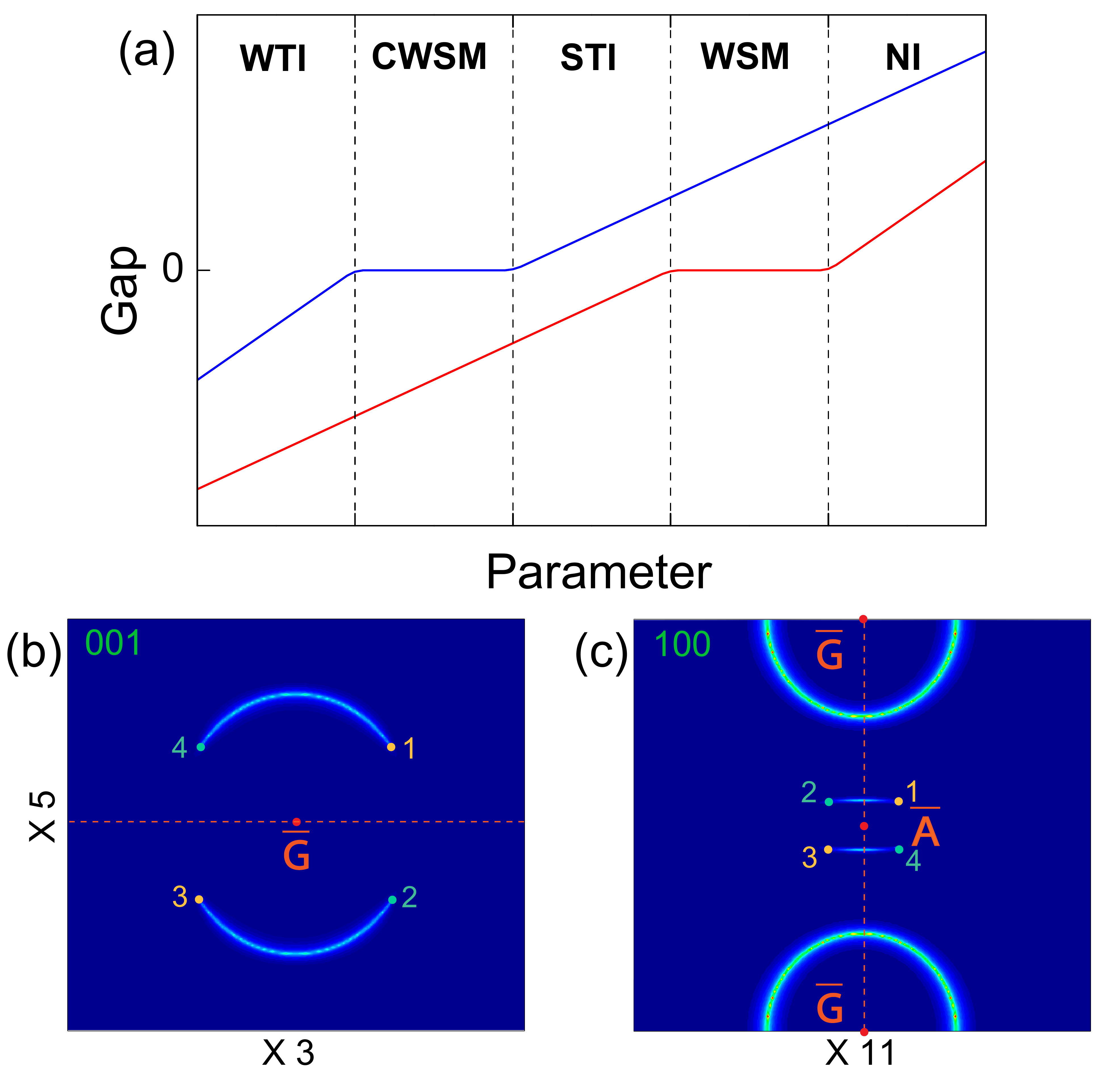}
\caption{{\bf Phase diagram and surface states for composite Weyl semimetal.} 
(a) Schematic phase diagram of WTI, CWSM, STI, WSM, and NI.
The two curves depict the direct gaps near the relevant $L$ and $M$ points.
(b) Existence of two open Fermi arcs at the $(001)$ surface of the calculated CWSM.
(c) Coexistence of two open Fermi arcs and one closed Fermi circle at the $(100)$ surface of the calculated CWSM.
In (b) and (c), the dimensions have been magnified for clarity.
}
\label{fig4}
\end{figure}

\noindent{\bf Composite Weyl Semimetals}

In the presence of both $\mathcal{P}$ and $\mathcal{T}$ symmetries, all bands are doubly degenerate and a Dirac point accounts for the NI-STI trainsition.
The Dirac point is unstable unless there is a symmetry protection~\cite{SYoung,XDai,SYang,BYang}.
However, when $\mathcal{P}$ or $\mathcal{T}$ symmetry is broken, the Dirac point splits into pairs of Weyl points, 
each of which is locally protected by the Chern number of a constant-energy surface enclosing it. 
One might naively think the same WSM phase~\cite{Volovik,Murakami,Wan,Burkov,Ran,Bernevig} 
naturally emerges near the WTI-STI transition.
In fact, the emergent phase turns out to be novel.
We now examine such a phase using $\beta$-Bi$_4$I$_4$ under $3.3\%$ compressive strain (see Fig.~\ref{fig3}a).
The $\mathcal{P}$ symmetry can be further broken by a possible energy difference between the $\rm Bi$ and $\rm Bi'$ atoms.
Because of the lack of a rotational symmetry, only two pairs of Weyl points appear at the WTI-STI transition; the bulk is a standard WSM.
At the $(001)$ surface, there are two open Fermi arcs connecting the four projected Weyl nodes, as shown in Fig.~\ref{fig4}b.
At the $(100)$ surface, surprisingly, there exists one closed Fermi circle in addition to the anticipated Fermi arcs, as shown in Fig.~\ref{fig4}c.
We dub this novel phase the {\em composite} WSM (CWSM).
When $\mathcal{P}$ asymmetry is small, while the Weyl points emerge near $L$, the band gap remains inverted at $M$.
As a consequence, there are surface states locally, separated by a large crystal momentum, 
at the $(100)$ surface where $L$ and $M$ project into distinct points $\bar A$ and $\bar G$. 
In contrast, at the $(001)$ surface where $L$ and $M$ project into the same point $\bar G$, 
the Fermi arcs appear attribute to the nontrivial Chern numbers, whereas the Fermi circle disappears because of its strong scattering with the metallic background.
Intriguingly for a CWSM, the Fermi arcs generically appear at any surface like the case for WSM, 
whereas the Fermi circle only appears in selected surfaces like the case for WTI.
In the most generic topological phase diagram, as sketched in Fig.~\ref{fig4}a,
the novel CWSM phase emerges near the WTI-STI transition when $\mathcal{P}$ or $\mathcal{T}$ symmetry is broken.\\

\noindent{\bf Discussions}

One may wonder whether $\beta$-Bi$_4$I$_4$ and $\beta$-Bi$_4$Br$_4$ are TCIs~\cite{Fu2012}, since a $(010)$ mirror symmetry is present. 
In a mirror invariant plane, the two spin subspaces, related by $\mathcal{T}$ symmetry, decouple and have opposite Chern numbers $\pm N$.
There must be $N$ pairs of protected Fermi points in the mirror invariant line (MIL) at the surface, independent of the Fermi level.
Similar to Bi$_2$Se$_3$~\cite{FZ13}, the STI $\beta$-Bi$_4$I$_4$ is also a TCI ($N=1$), 
because there are always two surface-state Fermi points in the MIL $\bar{G}\bar{X}$, as shown in Fig.~\ref{fig2}(c).
In contrast, the WTI $\beta$-Bi$_4$Br$_4$ is not a TCI ($N=0$), because Fermi points are not necessarily present along the MIL $\bar{G}\bar{A}$, 
as seen in Figs.~\ref{fig2}(f) and~\ref{fig2}(g).
These arguments are consistent with our direct calculations on the mirror Chern numbers~\cite{Teo2008}.

$\beta$-Bi$_4$X$_4$ (X=I,~Br) can be prototype WTI and offer a new platform for exploring exotic physics with simple chemistry.
In addition to the strain-induced topological phase transitions and CWSM phase,
the two $(100)$ surface states is particularly appealing, 
e.g., the spin texture, the Landau level crossing, and the exciton condensation triggered by the nesting of the electron and hole pockets.
In the thin-film QSHI limit, the small velocity along the $\bm a$ or $\bm c$ axis can be utilized to study 
the helical Luttinger liquid~\cite{RRD} and the $\mathcal{Z}_4$ parafermions~\cite{Z41,Z42,Z43}.
As for the CWSM, physics can be enriched by single-particle couplings or many-body interactions between the open Fermi arcs and the closed Fermi circle.

Finally, WTI have been proved to be {\em strong} against disorder~\cite{Stern12,Mong12,Kane12}, 
as long as $\mathcal{U}(1)$, $\mathcal{T}$, and translational symmetries are respected on average.
Breaking any symmetry at the $(100)$ surface of $\beta$-Bi$_4$Br$_4$ or strained $\beta$-Bi$_4$I$_4$ may induce an exotic phenomenon. 
A topological defect, e.g., a screw dislocation~\cite{Ran-WTI} or a step edge~\cite{Zhou15}, can break a translational symmetry binding a helical edge modes.
A Zeeman field can break $\mathcal{T}$ symmetry yielding a quantum anomalous Hall effect 
with a Chern number between $-2$ to $2$, which is tunable by the field orientation~\cite{FZ13}. 
A proximity coupling to an $s_{\pm}$ wave (e.g., iron-based) superconductor, can break $\mathcal{U}(1)$ gauge symmetry 
producing a $\mathcal{Z}_2$ topological superconductor with a Majorana Kramers pair~\cite{TSC13}.  \\

\noindent{\bf Computational Methods}

Our DFT calculations are performed using the projector augmented wave method implemented in the Vienna $\textit{ab initio}$ simulation package (VASP)~\cite{Kresse1996}.  
Perdew-Burke-Ernzerhof parametrization of the generalized gradient approximation (GGA-PBE) is used for the exchange correlation potential~\cite{Perdew1996,KJ99}. 
The plane wave energy cutoff is set to $300$~eV, and the Brillouin zone is sampled by a $9\times9\times6$ mesh. 
Based on the experimental or the optimized lattice structures, we apply the more sophisticated and accurate Heyd-Scuseria-Ernzerhof (HSE) hybrid functional method~\cite{Heyd2003} to the calculations of electronic band structures, 
using a $300$~eV wave energy cutoff and a $6\times6\times4$ Brillouin zone mesh. 

In studying the surface states, 
we first use the bulk DFT results and the Wannier90 code~\cite{Mostofi2008,Marzari1997,Souza2001} 
to construct the maximally localized Wannier functions for the $p$ orbitals of Bi and halogens.
Based on the Wannier functions, we then use an iterative method~\cite{Sancho1985} to obtain the surface Green's functions of semi-infinite systems, 
from which we calculate the dispersions of surface states.

Under strain, the atomic positions are allowed to be optimized 
until the force on every ion is less than $0.01$~eV/\AA~by employing the vdW corrections~\cite{Dion2004,Klimes2011}.
Since $\beta$-Bi$_4$Br$_4$ has yet to be synthesized in experiment, we calculate the phonon spectrum to investigate its lattice stability, using 
the PHONOPY code~\cite{Togo2008}. 
The vdW corrections and lattice relaxations are also employed to obtain the optimized atomic positions and lattice parameters for $\beta$-Bi$_4$Br$_4$,
which are list in Table S1~\cite{SM}.\\

\noindent{\bf Acknowledgments}

We thank the Kavli Institute for Theoretical Physics and the Aspen Center for Physics (05/2015-06/2015) for hospitality during the finalization of this work.


\begin{thebibliography}{100}
\bibitem{footnote} Because of the lack of $\mathcal{C}_3$ or $\mathcal{C}_4$ symmetry, only $\sigma_y$ and $\tau_z$, 
instead of all the $\bm\sigma$ and $\bm\tau$ operators, can be related to the electron spins and the orbital parities.
\end{thebibliography}
\end{document}